\documentclass[twocolumn,english]{revtex4-1}
\usepackage{lmodern}
\usepackage{lmodern}
\usepackage[T1]{fontenc}
\usepackage[latin9]{inputenc}
\usepackage{babel}
\usepackage{units}
\usepackage{amsmath}
\usepackage{amssymb}
\usepackage{graphicx}
\usepackage[unicode=true,pdfusetitle,
 bookmarks=true,bookmarksnumbered=false,bookmarksopen=false,
 breaklinks=false,pdfborder={0 0 1},backref=false,colorlinks=false]
 {hyperref}
\usepackage{breakurl}

\makeatletter

\usepackage{babel}
\usepackage{units}
\usepackage{breakurl}
\usepackage{units}

\usepackage{microtype}
\usepackage{mathbbol}

\begin{document}

\title{Current enhancement due to field-induced dark carrier multiplication in graphene}

\author{Roland Jago$^1$}
\author{Florian Wendler$^2$}
\email{florian.wendler@tu-berlin.de}
\author{Ermin Malic$^1$}
\affiliation{$^1$Chalmers University of Technology, Department of Physics, SE-412 96 Gothenburg, Sweden}
\affiliation{$^2$Institut f\"ur Theoretische Physik, Technische Universit\"at Berlin,  10623 Berlin, Germany}

\begin{abstract}
We present a microscopic study on current generation in graphene in response to an electric field. While scattering is generally considered to reduce the current, we reveal that in graphene Auger processes give rise to a current enhancement via a phenomenon we denote dark carrier multiplication. Based on a microscopic approach, we show that, if other scattering channels are absent, this prevents the carrier distribution to reach a stationary value. Taking into account scattering with phonons a finite current is restored, however its value exceeds the stationary current without scattering.

\end{abstract}
\maketitle

Transport properties of graphene, in particular the exceptionally
high electrical conductivity even at room temperature, have been intensively
studied since its discovery \cite{Geim2007,Morozov2008,CastroNeto2009,DasSarma2011Review}.
In early graphene samples, electrical transport was limited by disorder
resulting in mobilities of up to $\unit[20000]{cm^{2}V^{-1}s^{-1}}$
at low temperatures \cite{Novoselov2005,Zhang2005,Mariani2008}. However,
it was demonstrated that eliminating the extrinsic disorder, the fundamental
limit of the mobility at room-temperature is considerably higher
\cite{Morozov2008,Du2008,Bolotin2008}. Depending on the sample, flexural
or in-plane phonons are considered to be responsible for limiting
the intrinsic conductivity \cite{Katsnelson2008,Mariani2008,Mariani2010_PHONONS,Castro2010}.
Moreover, in contrast to conventional materials with a parabolic 
band structure, carrier-carrier scattering has an impact on the current
\cite{Kashuba2008,Fritz2008,Gornyi2012,Sun2012}. While most studies
in literature focus on the linear response and deploy the Drude approach
for the conductivity, there are only a few studies addressing the
non-linear response of graphene to an electric field \cite{Bistritzer2009,Balev2009,Dora2010,Rosenstein2010,Tani2012}.

In this work, we provide a microscopic access to the time- and momentum-dependent
carrier dynamics in graphene in a constant in-plane electric field. Using the density matrix formalism,
we calculate all intrinsic carrier-phonon and carrier-carrier scattering
channels within the second-order Born-Markov approximation. This allows us
to investigate the temporal evolution of the carrier density as well
as the generation and the dynamics of the electrical current giving
new insights into carrier transport in graphene. 
In a many-particle process that we denote as dark carrier multiplication (dark CM), Coulomb-induced processes bridging the valence and the conduction band (Auger scattering) significantly increase the carrier density in response to the electric field. 
Signatures of this effect have already been found in near-infrared transient absorption measurements under high electric fields using THz excitation pulses \cite{Tani2012}.
Furthermore, a carrier density increase has been discussed in literature as a consequence of the radiative coupling \cite{Balev2009} or the Schwinger mechanism \cite{Dora2010,Rosenstein2010}.
The latter two effects are much weaker at room temperature than
Auger processes, which have
been demonstrated to be extremely efficient in graphene 
\cite{Winzer2010_Multiplication, Winzer2012, Brida2013,Ploetzing2014,05_Mittendorff_Auger_NatPhys_2014,02_Wendler_CM_NatCommun_2014,Gierz2015}.
The aim of this work is to investigate the impact of the field-induced dark CM on the generation and enhancement of electric currents in graphene. 

\begin{figure}[!t]
\begin{centering}
\includegraphics[width=1\columnwidth]{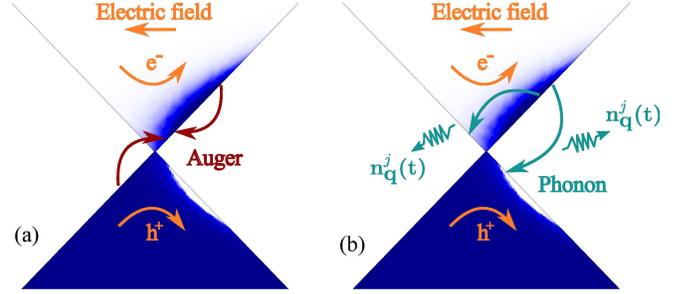} 
\par\end{centering}
\caption{Carrier occupation near the Dirac point in response to an in-plane electric
field. The blue filling represents the electron occupation, while
the white region in the valence band describes the hole occupation.
The electric field accelerates electrons and holes (orange arrows),
which opens up relaxation channels including Auger (red
arrows) and phonon-induced scattering (green arrows). }
\label{fig:sketch} 
\end{figure}

When graphene is placed in an external in-plane electric field $\mathbf{E}$,
its electrons are accelerated in the direction anti-parallel
to the field which we set to be the negative x-direction
$\mathbf{E}=(-E,0,0)$. This is illustrated in Fig. \ref{fig:sketch}
showing the Dirac cone of graphene including resulting Coulomb- and
phonon-induced scattering processes. 
Since we consider neutral graphene characterized by a vanishing chemical
potential, the electron-hole symmetry allows us to focus on the dynamics of electrons
in the conduction band. 
Nevertheless, it is instructive to note that
electrons and holes appearing on the same side of the Dirac cone,
move into opposite directions in real space, since the electron group velocity
is given by $\mathbf{v}_{\mathbf{k}}^{\lambda}=\lambda v_{\text{F}}\mathbf{k}/|\mathbf{k}|$
with the Fermi velocity $v_{\text{F}}$ and the band index $\lambda=\pm1$
denoting the conduction ($\lambda=+1$) and the valence ($\lambda=-1$)
band, respectively. Therefore, electrons at $\mathbf{k}$ move into
the $\lambda\mathbf{k}$-direction, and, 
since a hole is nothing else than a missing electron, it can be represented by an electron moving into the $-\lambda\mathbf{k}$-direction, which corresponds to a positively charged hole moving into the $\lambda\mathbf{k}$-direction.
Hence, the group velocities of electrons and holes are the
same, both changing sign when the band is switched. Consequently,
the shift of electron and hole occupations to the right in k space,
cf. Fig. \ref{fig:sketch}, means that electrons and holes move into opposite directions in real space.

\begin{figure}[!t]
\begin{centering}
\includegraphics[width=1\columnwidth]{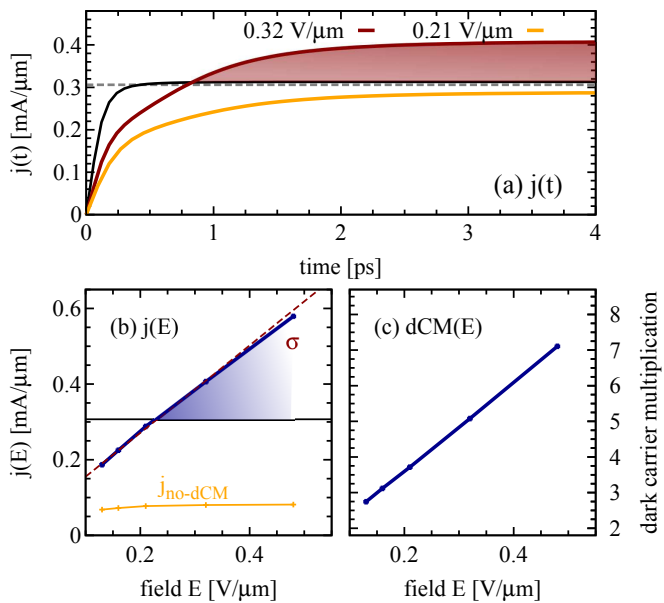} 
\par\end{centering}
\caption{(a) Dynamics of the current density $j(t)$ (in field direction) in graphene
subject to an in-plane dc electric field at room temperature. The
black line represents the result in the absence of many-particle
scattering processes and the dashed  gray line illustrates the saturation
current density given by Eq. (\ref{eq:j_max}). The red-shaded area shows the region, where current density enhancement takes place. (b) Equilibrium current density for
varying electric fields $E$ (blue line). Here, the dashed red line is a fit to the linear
region, where the slope corresponds to the conductivity $\sigma$. 
The black and yellow lines describe the
current density without taking into account any scattering processes or neglecting the dark carrier multiplication
(dCM), respectively. (c) Field-dependent dCM.}
\label{fig:current+CM(field)} 
\end{figure}

As electrons are accelerated in the electric field, the
magnitude of their velocity remains constant and only the direction
of their motion changes. Consequently, if scattering channels are
switched off, the current density saturates to a finite value corresponding
to the situation, in which all electrons move into the direction anti-parallel
to the field. The resulting current density dynamics exhibits an initial increase
until the equilibrium current density is reached after a few picoseconds,
cf. the black line in Fig. \ref{fig:current+CM(field)} (a). While
the time scale on which the saturation takes place is given by the
field strength $E$, the saturation current density only depends on the initial
temperature-dependent carrier density, cf. the black line in Fig.
\ref{fig:current+CM(field)}(b). 

The current density dynamics at room temperature
including all scattering channels is shown in Fig. \ref{fig:current+CM(field)}(a)
for different field strengths $E$. For $E<\unit[0.22]{V \mu m^{-1}}$,
the current density saturates, as expected, below the value without scattering
suggesting that a certain non-zero resistivity has been introduced.
Surprisingly, we reveal that for sufficiently high $E$ many-particle scattering
processes can even enhance the saturation current density, i.e. they introduce
a scattering-induced current density amplification, cf. the red-shaded area in Fig. \ref{fig:current+CM(field)}(a). This can be traced back to Auger scattering, which give rise to
a significant increase of the charge carrier density. This process bridging the valence and the conduction
band (Fig.\ref{fig:sketch}(a)) corresponds to the creation of an additional electron-hole pairs 
which indirectly -- via an enhanced carrier density -- boosts the
generated current density.
In analogy to the regular carrier multiplication (CM) induced by an
optical excitation \cite{Winzer2010_Multiplication}, we define the
field-induced dark CM by the ratio of the equilibrium carrier density
with ($n$) and without ($n_{0}$) the electric field: $\text{dCM}=n/n_{0}$.
The generated dark CM is shown in Fig. \ref{fig:current+CM(field)}(c) as a function of the electric field. Its increased efficiency explains the observed enhancement of the current density at high fields above the saturation value without scattering. 

Figure \ref{fig:current+CM(field)}(b) demonstrates that Ohm's law
${\bf j=\sigma{\bf E}}$ is valid in graphene for small fields, cf.
the linear increase with
a slope corresponding to a conductivity of $\sigma=\unit[1.1]{k\Omega^{-1}}$. This is larger than experimentally obtained values \cite{Novoselov2005,Zhang2005,Morozov2008}, since we consider a perfect graphene sample and an idealized situation without any negative influences of the environment. To illustrate that the large conductivity results from the strong impact of the dark CM, we approximate the current density
excluding the influence of the dark CM $j_{\text{no-dCM}}=j-j_{\text{dCM}}$,
cf. the yellow line in Fig. \ref{fig:current+CM(field)}(b). This
approximation is obtained using the relation $j=e_{0}n\, v_{\text{drift}}$ with the drift velocity $v_{\text{drift}}$
(average velocity of charged carriers). Now, the initial carrier density $n_{0}$ is separated from the dark
CM-induced density $n_{\text{dCM}}$ resulting in $j_{\text{no-dCM}}=e_{0}n_{0}\, v_{\text{drift}}$.
In the absence of dark CM, the current density is much smaller and shows only
a minimal increase with the field strength. Most importantly, it always
stays well below the saturation
current density in the case without scattering.

The mobility $\mu$ can be estimated via  the slope of the drift velocity $v_{\text{drift}}$
plotted over the electric field yielding in the linear region a value of $\mu\approx\unit[4,000]{cm^{2}V^{-1}s^{-1}}$.
It cannot be inferred from the usual relation $\mu=\sigma/(e_{0}n)$,
where $e_{0}$ is the elementary charge, since now the carrier density
$n$ depends on the electric field due to the appearance of dark CM. Instead,
using $j=e_{0}n\, v_{\text{drift}}$ and the generalized definitions
of the mobility and the conductivity $\mu=\text{d}v_{\text{drift}}/\text{d}E$
and $\sigma=\text{d}j/\text{d}E$, respectively, the relation $\mu=\sigma/(e_{0}n)-(v_{\text{drift}}/n)(\text{d}n/\text{d}E)$
is found. The reason for the comparably small value of the mobility (despite the rather large conductivity), is the relatively high temperature ($T=\unit[300]{K}$) corresponding to a large carrier density, which is even enhanced by the dark CM. 

Before we go further into detail of the carrier dynamics leading to
the enhancement of the current density via many-particle scattering, we first
briefly introduce the applied microscopic approach. The many-particle Hamilton
operator $H=H_{\text{0}}+H_{\text{c-c}}+H_{\text{c-ph}}+H_{\text{c-f}}$
consists of the (i) free carrier and phonon contribution $H_{\text{0}}$,
(ii) carrier-carrier $H_{\text{c-c}}$ and (iii) carrier-phonon $H_{\text{c-ph}}$
interaction accounting for Coulomb- and phonon-induced scattering,
and (iv) the interaction of an external electric field with carriers.
The latter contribution reads for electrons in second quantization
\cite{Meier1994} 
\begin{align}
H_{\text{c-f}} & =-ie_{0}\mathbf{E}\cdot\sum_{\mathbf{k}}a_{\mathbf{k}\lambda}^{\dagger}\nabla{}_{\mathbf{k}}a_{\mathbf{k}\lambda},\label{cfH}
\end{align}
where $a{}_{\mathbf{k}\lambda}^{\dagger}$, $a{}_{\mathbf{k}\lambda}$
denote creation and annihilation operators for electrons in the band
$\lambda$ and with the momentum \textbf{$\mathbf{k}$}. Evaluating the Heisenberg's equation
of motion and applying the second-order
Born-Markov approximation \cite{KochBuch,Kira2006PQE,MalicBuch}, we obtain the 
graphene Bloch equations explicitly including the impact of an electric field: 
\begin{align}
\dot{\rho}_{\mathbf{k}}^{\lambda}(t) & =\Gamma_{\mathbf{k}\lambda}^{\text{in}}\,\big(1-\rho_{\mathbf{k}}^{\lambda}\big)-\Gamma_{\mathbf{k}\lambda}^{\text{out}}\,\rho_{\mathbf{k}}^{\lambda}-\frac{e_{0}}{\hbar}\mathbf{E}\cdot\nabla_{\mathbf{k}}\rho_{\mathbf{k}}^{\lambda},\label{eq:rho}\\[10pt]
\dot{n}_{\mathbf{q}}^{j}(t) & =\Gamma_{\mathbf{q}j}^{\text{em}}\,\big(n_{\mathbf{q}}^{j}+1\big)-\Gamma_{\mathbf{q}j}^{\text{abs}}\, n_{\mathbf{q}}^{j}-\gamma_{\text{ph}}\,\big(n_{\mathbf{q}}^{j}-n_{\mathbf{q},\text{B}}^{j}\big).\label{eq:n}
\end{align}
The equations describe the coupled dynamics of the electron occupation probability
$\rho_{\mathbf{k}}^{\lambda}=\langle a_{\mathbf{k}\lambda}^{\dagger}a^{\phantom{\dagger}}_{\mathbf{k}\lambda}\rangle$
in the state $(\mathbf{k},\lambda)$ and the phonon number $n_{\mathbf{q}}^{j}=\langle b_{\mathbf{q}j}^{\dagger}b^{\phantom{\dagger}}_{\mathbf{q}j}\rangle$
in the mode $j$ with the momentum $\mathbf{q}$. The time- and momentum-dependent
in- and out-scattering rates $\Gamma_{\mathbf{k}\lambda}^{\text{in/out}}(t)$
contain all relevant carrier-carrier and carrier-phonon
scattering channels including acoustic $\Gamma\text{LA}$, $\Gamma\text{TA}$
and optical $\Gamma\text{LO}$, $\Gamma\text{TO}$, KLO, KTO phonons.
The dynamics of the phonon number $n_{\mathbf{q}}^{j}$ is determined
by the emission and absorption rates $\Gamma_{\mathbf{q}j}^{\text{em/abs}}(t)$
driving $n_{\mathbf{q}}^{j}$ towards the initial Bose-distribution
$n_{\mathbf{q},\text{B}}^{j}$. The  phonon
decay rate $\gamma_{\text{ph}}$ is adjusted to the
experimentally obtained value \cite{Kang2010}. For more details on the microscopic
origin of the rates, we refer to Refs. \onlinecite{MalicBuch, Malic2011}.

The presence of an electric field enters the Bloch equations through
the scalar product between the electrical field ${\bf E}$ and the
gradient $\nabla_{\mathbf{k}}$ of the carrier occupation $\rho_{\mathbf{k}}^{\lambda}$,
cf. Eq. (\ref{eq:rho}). It stems from
the commutation of the carrier-field Hamilton operator from Eq. (\ref{cfH})
with $\rho_{\mathbf{k}}^{\lambda}$. Within the applied Born approximation, the
electric field does not influence the lattice positions, hence there
is no electric field contribution in Eq. (\ref{eq:n}). Using the coordinate transformation
\cite{Meier1994} $\mathbf{k}\rightarrow\mathbf{k}-e_{0}\mathbf{E}\, t/\hbar$,
one can transform the Bloch equations according to $d/dt\rightarrow d/dt-e_{0}\mathbf{E}\cdot\nabla_{\mathbf{k}}/\hbar$ resulting in 
regular Bloch equations without the electric field. Here,
the field-induced dynamics is hidden in the new field- and time-dependent
momenta $\mathbf{k}-e_{0}\mathbf{E}\, t/\hbar$. This means that we consider
a moving coordinate frame describing the electric field-induced shift
of the carrier occupation $\rho_{\mathbf{k}}^{\lambda}$ in momentum
space.
 In case of a vanishing electric field $E=0$, the equilibrium distributions of electrons and phonons
$\rho_{\mathbf{k}}^{\lambda}$ and $n_{\mathbf{q}}^{j}$ are given
by Fermi-Dirac and Bose-Einstein distributions, respectively. When
this equilibrium is disturbed by a field $E\neq0$, many-particle
scattering occurs until new equilibrium distributions emerge. In conventional
semiconductors, the carrier densities of the two equilibrium distributions
(with and without the field) are the same. In contrast, strongly efficient
Auger scattering in graphene results in a carrier density increase,
which is quantified by the dark CM $n=\text{dCM}\cdot n_{0}$ \footnote{ Note that for an infinitely extended graphene sheet in an in-plane
electric field, electron-hole pairs can also be created due to the
Schwinger mechanism \cite{Schwinger1951,Dora2010,Rosenstein2010}.
Here, we restrict our investigation to finite temperatures
of $T\geq\unit[100]{K}$, where the field-induced acceleration of
thermal charge carriers gives rise to a dark CM that
prevails over the Schwinger effect. %
}. 

Numerical evaluation of the graphene Bloch equations provides a microscopic
access to the coupled time- and momentum-resolved carrier and phonon
dynamics under the influence of an electric field and its interplay
with Coulomb- and phonon-induced scattering channels. The main observable
in our study is the intraband current density $j(t)$ which reads under electron-hole symmetry \cite{MalicBuch} 
\begin{align}
\mathbf{j}(t) & =-\frac{8e_{0}}{L^{2}}\sum_{\mathbf{k}}\rho_{\mathbf{k}}^{\text{c}}(t)\frac{\nabla_{\mathbf{k}}\varepsilon_{\mathbf{k}}^{\text{c}}}{\hbar}=e_{0}n(t)\,\mathbf{v}_{\text{drift}}(t).\label{eq:j(t)}
\end{align}
The appearing area of the graphene flake $L^{2}$ cancels out after
performing the sum over all momenta ${\bf k}$. The factor 8 describes
the spin and valley degeneracy of the electronic states as well as the electron-hole symmetry. The current density
can be rephrased in terms of carrier density $n(t)=(8/L^{2})\sum_{\mathbf{k}}\rho_{\mathbf{k}}^{\text{c}}(t)$
and drift velocity $\mathbf{v}_{\text{drift}}(t)=-(8/L^{2}n(t))\sum_{\mathbf{k}}\rho_{\mathbf{k}}^{\text{c}}(t)\mathbf{v}_{\mathbf{k}}^{\text{c}}$,
where $\mathbf{v}_{\mathbf{k}}^{\text{c}}=\nabla_{\mathbf{k}}\varepsilon_{\mathbf{k}}^{\text{c}}/\hbar$
is the electron group velocity. In conventional materials with a parabolic band structure with the
effective mass $m$, the group velocity reads $\mathbf{v}_{\mathbf{k}}^{\text{c}}=\hbar\mathbf{k}/m$, whereas in graphene, it is given by
$\mathbf{v}_{\mathbf{k}}^{\text{c}}=v_{\text{F}}\mathbf{e}_{\mathbf{k}}$
due to its linear bandstructure $\varepsilon_{\mathbf{k}}^{\text{c}}=\hbar v_{\text{F}}|\mathbf{k}|$.
Consequently, only the angle between the electronic momentum $\mathbf{k}$ and the applied electrical field $\mathbf{E}$
determines the contribution of the carrier occupation $\rho_{\mathbf{k}}^{\text{c}}(t)$
to the current density in graphene, while the magnitude of $\mathbf{k}$ is
irrelevant in this regard. 

\begin{figure}[!t]
\begin{centering}
\includegraphics[width=1\columnwidth]{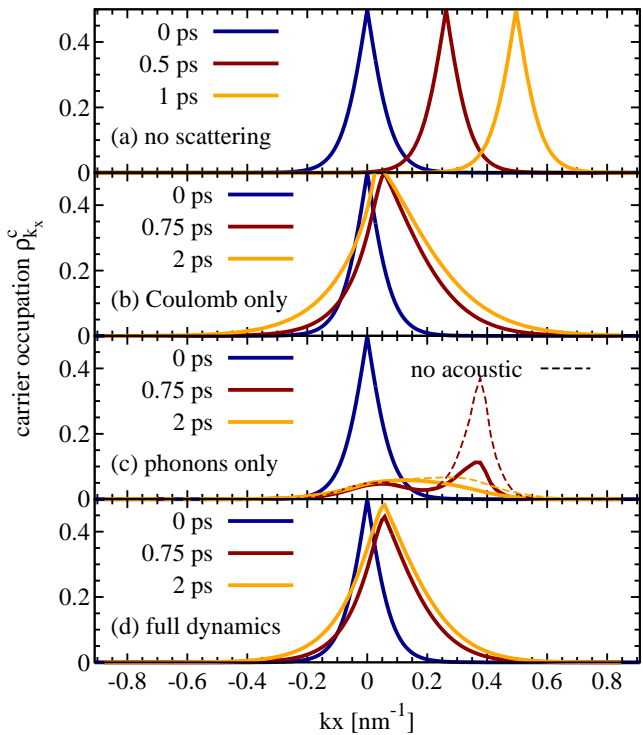} 
\par\end{centering}
\caption{Carrier occupation at different times
(with respect to the switch-on of the electric field) for a field strength of $E=\unit[0.32]{V\mu m^{-1}}$
and a temperature of $T=\unit[300]{K}$ including (a) no scattering
contributions, (b) only carrier-carrier scattering, (c) only carrier-phonon
scattering, and (d) full carrier dynamics.}
\label{fig:occupations} 
\end{figure}

The spectral distribution of the electron occupation $\rho_{\mathbf{k}}^{\text{c}}(t)$
is shown in Fig. \ref{fig:occupations} at room temperature and along
the $k_{x}$-direction (corresponding to the direction of the applied
electric field) at different times and including different scattering channels.
The case without scattering
demonstrates that the in-plane electric field accelerates the available
charge carriers in the k space into the opposite field direction ($-\mathbf{E}$),
cf. Fig. \ref{fig:occupations}(a). Initially without the field, the
carrier occupation is fully symmetric in $k$, hence the group velocities of the charge carriers point in different
directions resulting in zero drift velocity and current density, cf. Eq. (\ref{eq:j(t)}).
Switching on the electric field (at $t=0$), the carriers are shifted resulting in an asymmetric carrier distribution and thus a current density
is generated. It increases until the velocity of all charge carriers aligns with the field. 
The saturation current density sets in already at picosecond time scales and long before the charge carriers leave the linear region of the Brillouin zone, cf. the black line in Fig. \ref{fig:current+CM(field)} (a). 
The generated saturation current density is independent of the strength of the electric field (Fig. \ref{fig:current+CM(field)}(b)) and scales with $T^{2}$
\begin{align}
j_{\text{sat}}=\frac{8e_{0}v_{F}}{L^{2}}\sum_{\mathbf{k}}\rho_{\mathbf{k},0}^{\text{c}}=e_{0}v_{F}n_{0}=\frac{e_{0}\pi k_{B}^{2}}{3\,\hbar^{2}v_{F}}T^{2},\label{eq:j_max}
\end{align}
where $k_{B}$ is the Boltzmann constant. 

Next, we include all carrier-carrier scattering channels and observe that the
distribution is still slightly shifted with respect to the Dirac point and becomes spectrally broader
due to Auger scattering, cf. Fig. \ref{fig:occupations}(b). While
carrier-carrier scattering does not influence the current in conventional
materials, in graphene its impact on the current is twofold: (i) it
redistributes charge carriers and thereby induces a resistivity, i.e.
current reduction, and (ii) it results in a carrier density increase
via Auger processes (i.e. dark CM) giving rise to a current enhancement.
While mechanism (i) leads to a carrier distribution resembling a
Fermi distribution, which is shifted and distorted along the field
direction, mechanism (ii) causes a steady increase of the carrier
density without reaching an equilibrium. In contrast, including only
carrier-phonon scattering channels, cf. Fig. \ref{fig:occupations}(c),
an equilibrium carrier distribution is reached, however, it considerably deviates
from a Fermi distribution. 
While low energetic acoustic phonons play only a minor role for energy relaxation \cite{Malic2011,Winnerl2011},
they have a strong impact on momentum relaxation and thus on the current density, cf. dashed lines in Fig. \ref{fig:occupations}(c).
The complete dynamics including both carrier-carrier and carrier-phonon
scattering is displayed in Fig. \ref{fig:occupations}(d). Here, the
distribution resembles the case of purely Coulomb-induced dynamics. However, now it exhibits an equilibrium due to carrier-phonon
scattering competing with Auger processes and stabilizing the carrier density.

As a result, evaluating the graphene Bloch equations, the current density
dynamics is revealed on a microscopic footing. 
It results from the interplay of the field-induced acceleration of charge carriers and Coulomb- and phonon-induced scattering processes. 
We find current-reducing carrier-carrier and carrier-phonon scattering channels as well as current-enhancing Auger scattering, a specific Coulomb channel giving rise to a dark CM. It is small at low electric fields $E$, but causes a current density increase above the saturation value for $E>\unit[0.22]{V\mu m^{-1}}$, cf. Fig. \ref{fig:current+CM(field)}(b).

\begin{figure}[!t]
\begin{centering}
\includegraphics[width=1\columnwidth]{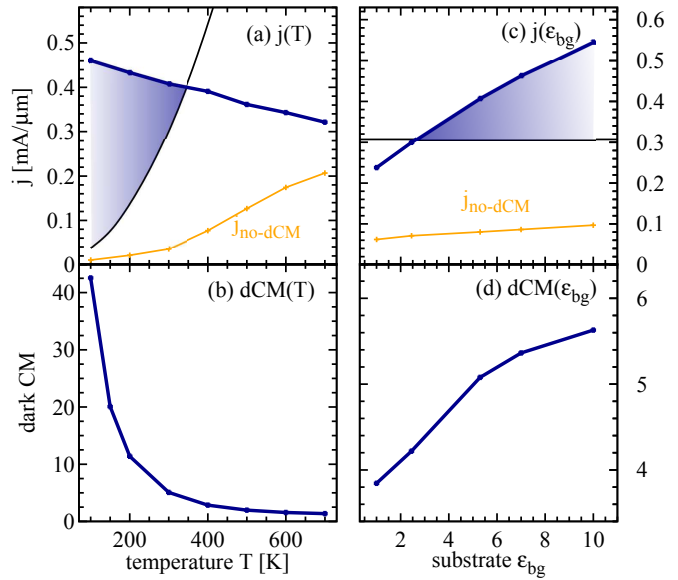} 
\par\end{centering}
\caption{Current density and dark carrier multiplication in dependence of
temperature $T$ and dielectric background constant $\varepsilon_{\text{bg}}$
with otherwise the same parameters as in Fig. \ref{fig:occupations}.
The thin black (yellow) lines show the behavior without many-particle scattering (excluding the impact of the dark CM). The blue-shaded areas illustrate the region, where current density enhancement takes place. }
\label{fig:current(temperature,substrate)} 
\end{figure}

Next, we investigate the dependence of the equilibrium current density
on the strength of the carrier phonon
scattering and the Coulomb interaction, which can be controlled by varying temperature and substrate, respectively.
The substrate is assumed to be only on one side of graphene
resulting in an averaged background dielectric constant $\varepsilon_{\text{bg}}=(\varepsilon_{\text{s}}+1)/2$.
The standard substrate used in Figs. \ref{fig:sketch}- \ref{fig:occupations}
is silicon carbide with a static dielectric constant $\varepsilon_{\text{s}}=9.66$ \cite{Patrick1970}. The crucial point in the temperature- and substrate-dependence of the current density is their influence on the dark CM, cf. blue and yellow lines in Fig. \ref{fig:current(temperature,substrate)}(a) and (c), respectively. 

The temperature dependence of the dark CM is determined by Pauli blocking, which is small for narrow carrier distributions at low temperatures, cf. Fig. \ref{fig:current(temperature,substrate)}(b). To understand the influence of the dark CM on the current density (Fig. \ref{fig:current(temperature,substrate)}(a)), we first consider the case without scattering (black line), where the current density is found to scale with $T^{2}$ according to Eq. (\ref{eq:j_max}). Switching on the scattering channels but suppressing the carrier density increase due to the dark CM (yellow line) the current density is strongly reduced, and owing to an enhanced acoustic phonon scattering this reduction increases with the temperature. Due to the pronounced dark CM at low $T$ (Fig. \ref{fig:current(temperature,substrate)}(b)), a significant current density amplification occurs up to the room temperature  (blue-shaded region).  For higher temperatures, the dark CM becomes negligible and the efficient scattering with acoustic phonons leads to a saturation of the current density.

The substrate dependence reveals a clear enhancement of the current density and the dark CM at higher dielectric background constants $\varepsilon_{bg}$, cf. Figs. \ref{fig:current(temperature,substrate)}(c) and (d). 
A larger $\varepsilon_{bg}$ screens the Coulomb potential and reduces the efficiency of carrier-carrier scattering. This leads to an enhanced dark CM, since the time window for Auger processes is increased. This resembles the increase of the conventional carrier multiplication at low pump fluences, cf. Ref. \onlinecite{Ploetzing2014}. Moreover, at low dielectric constants the Coulomb-induced redistribution of charge carriers competing with Auger scattering is suppressed resulting in improved conditions for the dark CM.

In conclusion, we have investigated the impact of the time- and momentum-resolved carrier dynamics on the generation of currents in graphene in the presence of an in-plane electric field. 
We show that field-induced acceleration of charge carriers provides
excellent conditions for Auger scattering, which give rise to a dark carrier multiplication resulting in a significant enhancement of the generated currents. 
The presented insights are applicably to the entire class of Dirac materials.

\begin{acknowledgments}
This project has received funding from the European Union's Horizon
2020 research and innovation programme under grant agreement No 696656 (Graphene Flagship).
Furthermore, we acknowledge support from the Swedish Research Council
(VR) and the Deutsche Forschungsgemeinschaft through SFB 658 and SPP 1459. The computations were performed on resources at Chalmers Centre for Computational Science and Engineering (C3SE) provided by the Swedish National Infrastructure for Computing (SNIC).
Finally, we thank Andreas Knorr (TU Berlin) for inspiring discussions. 
\end{acknowledgments}

\end{document}